\documentclass{PoS}

\title{
\rightline{\small IIT-HEP-10-6}
New Experiments with Antiprotons}
\ShortTitle{New Experiments with Antiprotons}

\author{\speaker{Daniel M. Kaplan}\thanks{For the Fermilab pbar Collaboration: Arizona: A.~Cronin; Bartoszek Engineering, L. Bartoszek; UC Riverside: A.\,P.~Mills, Jr.; Cassino: G.\,M.~Piacentino; Duke: T.\,J.~Phillips; FNAL: G.~Apollinari, D.\,R.~Broemmelsiek, B.\,C.~Brown, C.\,N.~Brown, D.\,C.~Christian, P.~Derwent, M.~Fischler, K.~Gollwitzer, A.~Hahn, V.~Papadimitriou, G.~Stancari, M.~Stancari, R.~Stefanski, J.~Volk, S.~Werkema, W.~Wester, H.\,B.~White, G.\,P.~Yeh; Ferrara: W.~Baldini; First Point Scientific: R.\,G.~Greaves; Hbar Tech.: S.\,D.~Howe, G.\,P.~Jackson, R.~Lewis, J.\,M.~Zlotnicki; Houston: K.~Lau; IIT Chicago: D.\,M.~Kaplan, T.\,J.~Roberts, J.~Terry, Y.~Torun, C.\,G.~White;  IIT Hyderabad, A.~Giri; ITEP Moscow: A.~Drutskoy; KSU: G.\,A.~Horton-Smith, B. Ratra; KyungPook: H.\,K.~Park; Lebedev: O.~Piskunova; Longwood: T.~Holmstrom; Luther Coll.: T.\,K.~Pedlar; Michigan: H.\,R.~Gustafson; Moscow State: M.~Merkin; NASA Marshall: J.\,B.~Pearson; Northwestern: J.~Rosen; Notre Dame: M.~Wayne; Oxford: F.~Azfar; SMU: T.~Coan; St.\ Xavier: A.~Chakravorty; Texas (Austin): M.\,G.~Raizen; Virginia: E.\,C.~Dukes.}  \thanks{Work supported by the U.S. Dept.\ of Energy.}%
      \\
       Illinois Institute of Technology\\
       E-mail: \email{kaplan@iit.edu} }

\abstract{Fermilab operates the world's most intense antiproton source. Newly proposed experiments can use those antiprotons either parasitically during Tevatron Collider running or after the Tevatron Collider finishes. For example, the annihilation of 8 GeV antiprotons  is expected to yield world-leading sensitivities to hyperon rare decays and {\em CP} violation. It could also provide  the world's most intense source of tagged $D^0$ mesons, and thus the best near-term opportunity to study charm mixing and, via {\em CP} violation, to search for new physics. Other precision measurements that could be made include properties of the $X(3872)$ and the charmonium system. An experiment using a Penning trap and an atom interferometer could make the world's first measurement of the gravitational force on antimatter. These and other potential measurements using antiprotons could lead to a broad physics program at Fermilab in the post-Tevatron era.}

\FullConference{35th International Conference of High Energy Physics\\
		 July 22-28, 2010\\
		 Paris, France}

\begin{document}

\section{Medium-Energy Antiproton Experiment}
Hyperon and charm {\em CP} violation provide windows into physics beyond the Standard Model that are complementary to those in the $B$ and $K$ sectors~\cite{Artuso-etal-Petrov,Tandean}. With substantial expected charm and hyperon production 
and low backgrounds, medium-energy antiproton annihilation can potentially yield event samples large enough to reveal  effects of new physics that may have been missed in previous studies.\footnote{The evidence for anomalous {\em CP} violation in $B_s$ mixing recently announced by the D\O\ collaboration~\protect\cite{D0-dimuon-anomaly}, if confirmed, may indicate that new physics does indeed contribute to {\em CP} asymmetries at detectable levels.} 
A very capable experiment for such studies can be mounted inexpensively at the Fermilab Antiproton Accumulator, by adding a magnetic spectro\-meter to the existing Fermilab E760 lead-glass calorimeter~\cite{CalNIM}, using an available BESS solenoid~\cite{BESS}, fine-pitch scintilla\-ting fibers (SciFi), the D\O\ SciFi readout system~\cite{D0-upgrade-NIM06}, and hadron ID via fast timing~\cite{psec}. If the relevant cross sections are as large as expected, this apparatus could enable world-leading studies of hyperon and charm {\em CP} violation and rare decays, charm mixing, and $X(3872)$ properties, along with those of other charmonium and nearby states, as is now proposed to Fermilab~\cite{P-986}. 
If approved, the experiment could start about a year after completion of the Tevatron Collider run.

\noindent {\bf Charm}~~ $D^0$--\,${\overline D}{}^0$ mixing is now established at $>$\,10 standard deviations~\cite{HFAG}, but the $\stackrel{<}{_\sim}$\,1\% values of the reduced mixing parameters  $x$ and $y$ are small enough to be possible Standard Model effects~\cite{Artuso-etal-Petrov}; the ``smoking gun'' for new physics in charm mixing will thus be {\em CP} violation~\cite{Artuso-etal-Petrov,Bigi-Uraltsev}. Unlike the beauty and kaon cases, Standard Model effects in charm are highly suppressed, offering a low-background venue for new-physics studies; moreover, some classes of new physics could be inherently more visible in interactions of ``up-type'' rather than ``down-type'' quarks. The  annihilation of 8\,GeV antiprotons with nucleons in a fixed target offers a charm cross section potentially of order microbarns, $10^3$ times larger than that at the B factories, along with low combinatoric background, making it possibly the ideal way to search for new physics via rare effects in charm.

\noindent {\bf Hyperons}~~ Having acquired (as of 1999) the world's largest HEP data sample, the Fermilab HyperCP experiment~\cite{Burnstein} found evidence (at $<$\,3 standard deviations) for new physics in hyperon {\em CP} violation and rare decays~\cite{Park,Materniak}, indicating the need for yet more data. An experiment designed to accomplish the charm studies just discussed can also acquire very large samples of hyperon decays and push precision hyperon studies to the next level of sensitivity.

\noindent {\bf X/Y/Z states}~~ The B factories  discovered new states in the charmonium region, not all of which can be charmonium levels~\cite{ELQ}. For example, the width of the puzzling $X(3872)$ resonance (seen in four experiments) may well be $\ll$\,1\,MeV~\cite{Braaten-Stapleton}. The unique ${\overline p}p$ precision, reflecting the small $\overline p$-beam $\delta p/p$ and absence of Fermi motion in a hydrogen target, is thus what is needed to establish whether this state is a $D^{*0}\overline{D}{}^0$ molecule~\cite{molecule}, a tetraquark state~\cite{Maiani}, or something else entirely.

\section{Antihydrogen Experiments}
\noindent{\bf In-flight {CPT} tests}~~ 
Production of antihydrogen in flight~\cite{Blanford}  may offer a way around some of the difficulties encountered in the CERN Antiproton Decelerator (AD)
trapping experiments. Methods have been proposed to measure in flight the antihydrogen Lamb shift and fine structure~\cite{Blanford-Lamb-shift}. Use of the Antiproton Accumulator for this purpose may be 
compatible with normal Tevatron Collider operations, and the program could continue into the post-Tevatron era. 

\noindent{\bf Antimatter Gravity Experiment}~~
While General Relativity predicts identical gravitational forces on matter and antimatter,  no direct experimental test has yet been made~\cite{Fischler-etal}. Effects of quantum  gravity,  
possible ``fifth forces,'' non-$1/r^2$ dependence, Lorentz violation 
\cite{Kostelecky}, etc.\  can show up in such  tests, 
now approved at the AD~\cite{AEGIS} and proposed at Fermilab~\cite{LoI}. The latter proposal  is to measure  a  slow ($\approx$\,1\,km/s) antihydrogen beam, formed in 
a Penning trap, using an atom interferometer composed either of material gratings (giving $\delta g/g\sim10^{-4}$) or  laser beams~\cite{Chu} ($\delta g/g\sim$\,$10^{-9}$). 
Deceleration for trapping can be accomplished down to $\approx$\,400\,MeV  in the Main Injector, 
followed by  an ``antiproton refrigerator''~\cite{refrigerator}, reverse linac, or  small synchrotron~\cite{LoI}.



\end{document}